\documentclass[twocolumn,aps,showpacs]{revtex4}
\usepackage{dcolumn}
\usepackage{bm}
\usepackage{amsthm}
\usepackage{color}

\usepackage{graphicx,epsfig}
\usepackage[english]{babel}
\usepackage{amssymb,amsfonts,amsmath}
\usepackage{verbatim}

\newcommand{\vev}[1]{\left\langle{#1}\right\rangle}
\def\bb#1{{\pmb{#1}}}
\def\d{\partial}

\def\e{\epsilon}
\def\bb#1{{\pmb{#1}}}
\def\d{\partial}

\def\e{\epsilon}
\def\-{\!-\!}
\def\+{\!+\!}

\begin{document}

\title{Skyrmions and Hall Transport}

\author{Bom Soo Kim}
\affiliation{Department of Physics and Astronomy, University of Kentucky, 
Lexington, KY 40506, USA }

\author{Alfred D. Shapere}
\affiliation{Department of Physics and Astronomy, University of Kentucky, 
Lexington, KY 40506, USA }
\date{\today}

\begin{abstract}
We derive a generalized set of Ward identities that captures the effects of topological charge on Hall transport. The Ward identities follow from the 2+1 dimensional momentum algebra, which includes a central extension proportional to the topological charge density. In the presence of topological objects like Skyrmions, we observe that the central term leads to a direct relation between the thermal Hall conductivity and the topological charge density. We extend this relation to incorporate the effects of a magnetic field and an electric current. The topological charge density produces a distinct signature in the electric Hall conductivity, which is identified in existing experimental data, and yields further novel predictions. For insulating materials with translation invariance, the Hall viscosity can be directly determined from the Skyrmion density and the thermal Hall conductivity to be measured as a function of momentum.
\end{abstract}

\maketitle

{\it Introduction:} 
Ward identities in quantum field theories \cite{Ward} are 
relations among correlation functions that follow solely from conservation equations,
and are thus independent of the properties of the Hamiltonian, other than its symmetries. 
Among their many applications in quantum field theory and many body physics (see for example 
\cite{Takahashi,Superconductivity,timedep}), Ward identities can be used to derive nontrivial relations among various measurable quantities such as conductivites and viscosities.
Recently, Ward identities have been applied to 2+1 dimensional systems with broken parity to 
show that the Hall viscosity is equal to one-half of the angular momentum in the presence of a gap, 
along with other relations involving transport coefficients  
\cite{Read:2008rn,Read:2010epa,Hoyos:2011ez,Bradlyn:2012ea,Geracie:2014nka,Hoyos:2014lla,Hoyos:2015yna}. 
In this letter, we obtain a more general and powerful set of Ward identities for such systems
by incorporating topological charges \cite{Watanabe:2014pea} 
that are not captured by conservation equations.  
These identities lead to useful relations between transport properties and 
the density of topological objects such as Skyrmions, with corresponding experimental signatures.   

Skyrmions \cite{Skyrmion,SkyrmionBaby,FracSpin} in magnetic materials are stable, particle-like spin textures that are protected by topological quantum numbers.
They have been studied theoretically 
\cite{SkyrmionTheory1,SkyrmionTheory2,SkyrmionTheory3,SkyrmionTheory4,SkyrmionTheory5,SkyrmionTheory6,SkyrmionTheory7}
and have been realized experimentally in magnetic materials 
\cite{SkyrmionExp1,SkyrmionExp2,SkyrmionExp3, Review}. 
Their transport properties have been measured, including their electric Hall conductivity 
\cite{HC1,HC2,HC3,HC4,HC5,HC6}, thermal Hall conductivity, and angular momentum 
\cite{SpinTorque1,ThermalHC1}. 
The identification of these quantities is often subtle, due to the Skyrmions' extended nature and 
their interactions with conduction electrons and other backgrounds, 
and has been based on phenomenological models rather than first principles. 
A better theoretical understanding of the transport properties of Skyrmions and 
their various relationships could be of great help in interpreting these experiments. 

We will derive from the advertised Ward identities a set of simple and universal relations 
among the observables mentioned above, in the presence of baby Skyrmions (Skyrmions in 2+1 dimensions), which we hereafter refer to simply as Skyrmions. 
In particular, we will argue that in insulators with translation and rotation invariance, 
the thermal Hall conductivity is proportional 
to the topological charge density, as in eq. \eqref{HallWI}.  
In the absence of translation invariance, these two quantities appear in a relation \eqref{WIInsulator} 
which also involves the Hall viscosity and angular momentum. 
According to our Ward identity, the Hall viscosity, which has been previously overlooked in both theoretical and experimental studies of Skyrmion systems, 
can be directly determined from the Skyrmion density and the thermal Hall conductivity measured as a function of momentum, as explained in eq. \eqref{MeasuringHV}.  

For metallic materials with Skyrmions, the tight binding between Skyrmion spins and the spins of conduction electrons
implies that the Skyrmion density makes a contribution to the electric Hall conductivity \cite{HC1,HC2,HC3,HC4,HC5,HC6}.  
We identify this contribution in our context in eq. \eqref{bBHallWI}, along with 
new experimental implications that should be possible to verify without difficulty. 
An expression for the Hall viscosity in terms of the measured electric Hall conductivity as a function of momentum is given in \eqref{MeasuringHV2}. 
Skyrmion dynamics turns out to be robust against the presence of impurities \cite{Impurity};
thus our conclusions should apply to realistic materials. 
\\

{\it Ward identities and central extension:} 
Topologically non-trivial objects can lead to significant modifications in certain physical quantities. 
Baby Skyrmions carry topological charge  that modifies the commutators of momentum operators 
\cite{Watanabe:2014pea,Toma1991,Thouless:1996}
\begin{align}\label{momentumalgebra}
[P^ i (x^0), P^j (x^0)] = i \hbar C^{ij} \;,
\end{align}  
where $x^0$ is time, $ x^\mu = (x^0,\vec x)$, and $i, j=1,2$. 
$P^i (x^0) \equiv \int d^2 x~ T^{0i} (x^\mu)$ is the momentum 2-vector, defined as the space integral of 
the momentum density components of the stress energy tensor $T^{\mu\nu}$. 
The central term on the right side of (\ref{momentumalgebra}) is proportional to the net topological charge $C^{ij}$, which may be written as the spatial integral of the topological charge density $C^{ij}= \int d^2 x~ c^{ij} (x^\mu)$. 
In terms of the spin configuration $\vec n(x^\mu)$,  
\begin{align}
c^{ij} = \vec n \cdot [\partial_i \vec n \times \partial_j \vec n] \;. 
\end{align}
We assume that the spin $\vec n$ varies smoothly over space and 
that the continuum description is valid on length scales much larger than the lattice spacing \cite{Review}. 
Since $c^{ij}$ is antisymmetric, we can write $c^{ij}\equiv c\, \epsilon^{ij}$. 
The modified commutator \eqref{momentumalgebra} implies that one cannot fully specify the momentum of the object due to uncertainty relations among different components. It is strongly reminiscent of the momentum algebra in the presence of a background magnetic field. 
As we will explain below, $c^{ij}$ can be interpreted as an effective magnetic field produced by the Skyrmions.   

A local form of the momentum commutator \eqref{momentumalgebra} is more convenient for our purposes. 
As proposed in \cite{Watanabe:2014pea},
\begin{equation} \label{CommRel}
\begin{split}
&[T^{0i}(x^0, \vec x), T^{0j}(x^0, \vec x')] \\
&= i \left( -\partial_i T^{0j}(x^\mu) 
+ \partial_j T^{0i}(x^\mu)+ c^{ij}  \right) \delta^{2} (\vec x-\vec x') \;,
\end{split}
\end{equation}
where the momentum density operators $T^{0i}(x^\mu)$ produce the derivative terms on the right side.  
Here and henceforth we set $\hbar=1$. 

We now proceed to generalize the Ward identities for systems with broken parity  
\cite{Read:2008rn,Read:2010epa,Hoyos:2011ez,Bradlyn:2012ea,Geracie:2014nka,Hoyos:2014lla,Hoyos:2015yna}
by including the central extension in the equal time commutator.   
For simplicity, time and space translation symmetries are assumed along with rotational symmetry 
in the spatial plane, which is compatible with the central extension in eq. \eqref{CommRel} in $2+1$ dimensions \cite{Watanabe:2014pea}. 
We begin by considering the retarded correlator of momentum densities  
\begin{equation}\label{RetardedCorr}
G^{0i,0j}(x^\mu, x'^\mu) \equiv i \theta(x^0-x'^0) \langle [T^{0i}(x^\mu), T^{0j} (x'^\mu)] \rangle . 
\end{equation}
Applying two time derivatives $\partial_0 \partial'_{0}$ to this expression produces four terms. 
When both derivatives act on the momentum density, we use local momentum conservation 
$ \partial_\mu T^{\mu i}=0$ %
\footnote{
Although translation symmetry is spontaneously broken due to the presence of Skyrmions, 
the conservation equations can be checked to hold with the model in \cite{Watanabe:2014pea}. 
Translation symmetry of the Skyrmion configuration can be restored in the continuum limit. 
} 
to obtain a term with two spatial derivatives of the retarded correlator, $\partial_n \partial'_{m} G^{ni,mj}$. 
All other terms contain a delta function $\delta (x^0-x'^0) $, 
the derivative of the step function in \eqref{RetardedCorr}. 
These latter terms are precisely the contact terms that have been reported to be missing 
in some evaluations of Kubo formulas \cite{Bradlyn:2012ea}. We stress that 
our Ward identities automatically produce all possible contact terms. 
For translation invariant systems, all the contact terms vanish \cite{Hoyos:2015yna} 
and the Ward identity becomes, in momentum space,   
\begin{equation}\label{MomWI}
\begin{split}
\omega^2 \widetilde G^{0i,0j}=
-i\omega \epsilon^{ij} c + q_m q_n \widetilde G^{ni,mj}\;,
\end{split}
\end{equation}
where $i, j, m, n$ are spatial indices, and a Fourier transform 
has been applied in the form 
$
\widetilde{G}^{0i,0j}(q_\mu)
\equiv \int  dx^0 d^2 x~ e^{i q_\mu x^\mu }  G^{0i,0j} (x^\mu), 
$
with $q_0\equiv \omega$.   
We may treat the Skyrmion density $c$ as a constant when discussing transport measurements on distance scales much larger than that of an individual Skyrmion. 

Rotation invariance constrains the components of the retarded Green's function. In general, 
$ \widetilde G^{ni,mj}$ can have 
three independent contributions: a shear viscosity term $-i\omega \eta (\delta^{nm}\delta^{ij}+\delta^{nj}\delta^{im}-\delta^{ni}\delta^{mj}) $, a bulk viscosity term $-i\omega \zeta\delta^{ni}\delta^{mj} $, and a Hall viscosity term  
$-\frac{i}{2} \omega \eta_H  (\epsilon^{nm}\delta^{ij}+\epsilon^{nj}\delta^{im}+\epsilon^{im}\delta^{nj} +\epsilon^{ij}\delta^{nm}) $ \cite{Avron:1995}. 
The coefficients $\eta$, $\zeta$, and $\eta_H$ can be complex functions of the frequency, whose real parts 
are the usual transport coefficients. The components with two spatial indices $\widetilde G^{0i,0j}$ take the form 
\begin{align}
-i\omega \left[ \delta^{ij} \bb{\kappa}_\delta 
\!+ \epsilon^{ij} \bb{\kappa}_\epsilon \!+ q^i q^j \bb{\kappa}_{q} 
\!+ (q^i \epsilon^{jn} \!+ q^j \epsilon^{in}) q_n \bb{\kappa}_{q \epsilon} \right],
\end{align}
where the form factors $\bb{\kappa}$ are the (complex) thermal conductivities, including the Hall component $\bb{\kappa}_\e$. These form factors are analogous to the electric conductivities coming from a 
retarded current-current  correlator that will be introduced later.  
By using these expressions to decompose equation \eqref{MomWI} into
independent tensor structures,
one obtains 
\begin{equation} \label{MomWI2}
\begin{split}
&\omega^2 \left[ \delta^{jl} \bb{\kappa}_\delta 
+ \e^{jl} \bb{\kappa}_\e
+ q^j q^l   \bb{\kappa}_{q}  + (\e^{jo}q^l \!+\! \e^{lo}q^j) q_o  \bb{\kappa}_{q\e}  \right] \\
&\qquad = \e^{jl}  \big[c  + q^2 \eta_H \big]  
+ \delta^{jl} q^2  \eta + q^j q^l  \zeta   \;, 
\end{split}
\end{equation}
which contains four distinct Ward identities corresponding to four independent tensor structures \cite{Hoyos:2015yna}.

By isolating the momentum independent terms in \eqref{MomWI2} proportional to $\delta^{jl}$ and $\epsilon^{jl}$, we arrive at the simple relations 
\begin{equation}\label{HallWI}
\begin{split}
& \omega^2 \bb{\kappa}^{(0)}_\delta = 0   \;, \qquad 
\omega^2 \bb{\kappa}^{(0)}_\epsilon = c  \;,
\end{split}
\end{equation} 
where the superscript $^{(0)}$ denotes the momentum independent part.  
Intuitively, the reason $\bb{\kappa}^{(0)}_\delta $ vanishes and $\bb{\kappa}^{(0)}_\epsilon $ does not is that   
Skyrmions are associated with spontaneously broken translation symmetry 
along with broken parity, whose imprints can only enter through the parity odd part of the 
conductivity at zero momentum. 
More precisely, the second identity predicts that the formation of a single Skyrmion results in the creation of 
a unit of thermal Hall conductivity $\bb{\kappa}^{(0)}_\epsilon$ 
in units of the quantized topological charge density. 
The frequency dependence is a consequence of the pole structure of the Goldstone boson that 
manifests itself in the retarded momentum correlator. 
In the presence of disorder, the behavior $ \bb{\kappa}^{(0)}_\epsilon = c /\omega^2 $ could in principle be lifted. 
However, recent numerical simulations have confirmed that Skyrmion motions are unaffected by impurities, 
in contrast to the case of domain walls \cite{Impurity}. 
The thermal Hall conductivity $\bb{\kappa}_\epsilon$ is dissipationless and exists even at zero temperature. 
While our Ward identity relations are valid at finite temperatures as well, measurements will be cleaner 
at very low temperatures, where additional dissipative contributions are suppressed. 
Another interpretation of eq. \eqref{HallWI} is that the Skyrmions carrying the thermal current propagate in an effective magnetic field given by the Skyrmion charge density $c_{ij}$, leading to a thermal Hall effect \cite{stone:1995}.  

For the momentum dependent terms in \eqref{MomWI2}, we obtain 
\begin{equation}\label{NeutralWIMomen}
\begin{split}
\omega^2 \bar{\bb{\kappa}}_\delta  
= q^2 \eta  \;, \quad
\omega^2 \bar{\bb{\kappa}}_\epsilon 
= q^2 \eta_H \;, \quad 
\omega^2 \bb{\kappa}_{q}  
= \zeta \;,
\end{split}
\end{equation} 
where the bar $~\bar{} ~$ indicates the nonconstant momentum dependent part; for example, 
$\bar{\bb{\kappa}}_\e = \bb{\kappa}_\e - \bb{\kappa}^{(0)}_\e = q^2 \bb{\kappa}_\e^{(2)} + q^4 \bb{\kappa}_\e^{(4)} + \cdots $. Thus, thermal conductivities are directly connected to the viscosities of the system, as previously confirmed \cite{Hoyos:2015yna}. 
Furthermore, it follows from \eqref{MomWI2} that $ \bb{\kappa}_{q\e} =0$. 

If the system of interest is not translationally invariant, there will be additional contributions 
to the Ward identity \eqref{NeutralWIMomen}; however, the zero momentum identity \eqref{HallWI} will be unmodified. 
A particularly interesting contribution of this type arises in parity-breaking systems exhibiting spontaneously generated angular momentum $\ell$ \cite{Toma1991,Liu:2012zm}, where 
the momentum generator can develop an expectation value 
\begin{equation}\label{Angmom}
\vev{T^{0i}}=\frac{1}{2}\e^{ik}\partial_k\ell \;.
\end{equation}
In the absence of translation invariance, the two time derivatives $\partial_0 \partial'_{0}$ acting 
on $G^{0i,0j}(x^\mu, x'^\mu)$ pick up the contact term 
$\frac{i}{2}(\partial'_0 - \partial_0) \big[ \delta(x^0\!- x'^0) \langle [T^{0i}(x^\mu), T^{0j}(x'^\mu)] \rangle \big]$ in addition to the terms appearing in \eqref{MomWI}. 
The commutator yields a tensor similar to $\eta_H$ coming from the last term in \eqref{MomWI}, as one can check using \eqref{CommRel} and \eqref{Angmom} \cite{Hoyos:2015yna}. 
As a result, $\eta_H$ in \eqref{NeutralWIMomen} is modified to $\eta_H  + \frac{\ell}{2}$. 
In such cases, a  coordinate space description  might be more convenient.  
Similarly, the inclusion of pressure $p$, another universal contribution, would replace $\zeta$ in the last relation 
of \eqref{NeutralWIMomen} by the combination $ \zeta -\frac{i}{\omega} p $.

{\it Ward identities for insulators:} 
Recently, Skyrmions have been observed in the insulating material Cu$_2$OSeO$_3$ 
\cite{Insulating1}, 
and various experiments regarding the Hall thermal conductivity and angular momentum have been carried out  
\cite{Rotation1,ThermalHC2,ThermalHC1}.
For insulators, our Ward identity provides a simple relation among parity violating transport 
coefficients,  
\begin{align}\label{WIInsulator}
&\omega^2 \bb{\kappa}_\e
= c - \partial^2 \Big( \eta_H + \frac{\ell}{2} \Big)  \;,
\end{align} 
which is derived from eqs. \eqref{MomWI2} and \eqref{Angmom} in the absence of translation invariance. 
Recent experiments have successfully measured the Skyrmion density, thermal Hall conductivity and angular momentum in Skyrmion materials \cite{ThermalHC1}. 
Such measurements could in principle be used to infer the existence of Hall viscosity 
\footnote{In real materials, $\bb{\kappa}_\e$ has a further contribution from the magnon \cite{Insulating1}, 
whose dynamics has been considered recently \cite{MagnonScattering}. 
Modeling the interaction between the Skyrmion and the magnon in this context of Ward identity 
is an interesting problem, which is beyond the scope of this letter due to the non-conservation of the magnon spin and the dependence on details of the interaction.}.

In the presence of translation invariance, there is a simple way to measure the Hall viscosity. Combining eqs. \eqref{HallWI} and \eqref{NeutralWIMomen}, we get 
\begin{equation} \label{MeasuringHV}
\begin{split}
\eta_H = c \frac{ \bar{\bb{\kappa}}_\epsilon}{q^2 \bb{\kappa}^{(0)}_\epsilon} 
\to c \frac{ \bb{\kappa}_\epsilon^{(2)} }{\bb{\kappa}^{(0)}_\epsilon}    \;,
\end{split}
\end{equation} 
where we take the limit $q^2 \to 0$. Once the thermal Hall conductivity $\bb{\kappa}_\epsilon$ is measured as a function of $q^2$, the Hall viscosity is nothing but the Skyrmion density multiplied by the ratio between the slope and $\bb{\kappa}_\epsilon$-intercept $\bb{\kappa}_\epsilon(q^2=0)$. Note that this is only applicable in the presence of nonzero Skyrmion density. 

{\it Ward identities for conductors:}  
The neutral case discussed above provides a simple relation between 
the topological charge density and thermal Hall conductivity. 
However, since most realistic materials reveal Skyrmions in the presence of electric charge carriers,
we need to generalize our discussion to include conducting materials. 
We will see that Skyrmions have a direct effect on charged dynamics as well, which can be 
accounted for by the inclusion of a conserved $U(1)$ current $J^\mu$, $\partial_\mu J^\mu =0 $. 

In the presence of a uniform external magnetic field $B$, the momentum density gets modified to 
\begin{align}\label{TB}
&T_B^{0 j} = T^{0 j} -(B/2) \epsilon^j_{\ k} x^k J^0 \;. 
\end{align} 
This modification is the expected minimal coupling in the presence of a constant magnetic field.
Another important modification appears in the conservation equation 
\begin{align}\label{EqB}
\partial_\mu T^{\mu i} = B \epsilon^i_{\ j} J^j \;. 
\end{align}
These are the spatial components of the general relation $\partial_\mu T^{\mu \nu} = F^{\nu\rho} J_\rho $.
The Ward identities are once again obtained by taking time derivatives of the correlator \eqref{RetardedCorr}.
The derivation is straightforward, and 
we present the details in the Supplemental Material. 
The resulting full Ward identity is 
\begin{align}\label{BMomWIFullMain}
& \delta^{jl}  [\omega^2 \bb{\kappa}_\delta 
\!+\! i\omega B \left(\bb{\alpha}_\e \!+\! {\bb{\alpha}}_\e^* \!+\! q^2 [ \bb{\alpha}_{q\e}\!-\! {\bb{\alpha}}_{q\e}^*]  \right) \!+\! B^2 \left( \bb{\sigma}_\delta \!+\! q^2 \bb{\sigma}_{q} \right) ] \nonumber \\
&\quad \!+\! \e^{jl} [\omega^2 \bb{\kappa}_\e
\!-\! i\omega B (\bb{\alpha}_\delta \!+\! {\bb{\alpha}}_\delta^* \!+\! q^2 [\bb{\alpha}_{q}\!+\!{\bb{\alpha}}_{q}^*]/2  ) \!+\! B^2 \bb{\sigma}_\e  ] \nonumber \\
&\!+\! q^j q^l  [ 
\omega^2 \bb{\kappa}_{q} \!-\! 2i\omega B (\bb{\alpha}_{q\e} \!-\! {\bb{\alpha}}_{q\e}^*) \!-\! 
B^2 \bb{\sigma}_{q} ]  \nonumber \\
&\quad
\!+\! (\e^{jo}q^l \!+\! \e^{lo}q^j) q_o [ 
\omega^2 \bb{\kappa}_{q\e} \!+\! i\omega B (\bb{\alpha}_{q} \!-\! {\bb{\alpha}}_{q}^*)/2 \!-\! B^2 \bb{\sigma}_{q\e} ] \nonumber \\
&= \e^{jl}  \big[c - B\rho + q^2 \eta_H \big]  
+ \delta^{jl} q^2  \eta + q^j q^l  \zeta   \;, 
\end{align}
where $\bb{\alpha}, {\bb{\alpha}}^*$ are thermoelectric conductivity tensors 
related to the form factors of momentum-current correlators
$G^{0i,j} \sim \langle [T^{0i}, J^{j}] \rangle $ and 
$G^{i,0j} \sim \langle [J^{i}, T^{0j}] \rangle $, while the
$\bb{\sigma}$'s are electric conductivity tensors associated with current-current correlators 
$G^{i,j} \sim \langle [J^{i}, J^{j}] \rangle $. 
They arise due to the modifications in equations \eqref{TB} and \eqref{EqB} and the corresponding mix between 
the momentum $T^{0i}$ and charge $J^{j}$ densities. 
There are four independent tensor structures and four corresponding Ward identities 
in \eqref{BMomWIFullMain}. These identities reduce to those of insulators when $B=0$, eq. \eqref{MomWI2}.

The momentum independent Ward identities give 
\begin{equation}\label{BHallWI}
\begin{split}
& \omega^2 \bb{\kappa}^{(0)}_\delta
+i\omega B (\bb{\alpha}^{(0)}_\e + {\bb{\alpha}}^{*(0)}_\e) + B^2 \bb{\sigma}^{(0)}_\delta  
= 0 \;, \\
& \omega^2 \bb{\kappa}^{(0)}_\e
-i\omega B (\bb{\alpha}^{(0)}_\delta + {\bb{\alpha}}^{*(0)}_\delta) + B^2 \bb{\sigma}^{(0)}_\e  
= c - B\rho \;,
\end{split}
\end{equation}
which reduce to eq. \eqref{HallWI} when $B=0$. 
At non-zero momentum, there are four independent relations connecting viscosities and  conductivities as in the neutral case. 
In particular, the Hall viscosity \cite{Avron:1995} is 
\begin{align} \label{BWIHallViscosity}
q^2 \eta_H = \omega^2 \bar{\bb{\kappa}}_\e + B^2 \bar{\bb{\sigma}}_\e 
-i\omega B \Big[\bar{\bb{\alpha}}_\delta + \bar{{\bb{\alpha}}}^*_\delta
+\frac{\bb{\alpha}_{q} + {\bb{\alpha}}_{q}^*}{2} \Big].
\end{align}

{\it Ward identities for conductors at zero momentum}: 
In \cite{FerroCoupling,FerroCoupling2,Review}  
interactions between Skyrmions and conduction electrons are modeled 
by the ferromagnetic spin coupling. 
In the strong coupling limit, the spin wave function of the conduction electrons is identified 
with that of the localized spin $\vec n(x^\mu)$ of the Skyrmions.  
This limit is described by a tight binding model with Hund's rule coupling. 
More general  interactions between conduction electrons and local magnetization may be 
considered \cite{FerroCoupling3}. 

We will discuss two different ways to model the effects of the interaction between the thermal 
and charge responses.  First, we can modify the parameters of the Ward identities. 
The Skyrmion charge density produces an emergent magnetic field $b=c/2$ \cite{Review}, 
which can change the dynamics of conduction electrons, similarly to $B$. For simplicity, we assume that
the emergent magnetic field is homogeneous and constant, which is the case for all practical measurements. 
Due to the tight binding, the motion of the 
conduction electrons will also influence the thermal response of the Skyrmions. 
At vanishing momentum, by taking these effects into account, we get 
\begin{equation}\label{bBHallWI}
\begin{split}
\omega^2 \bb{\kappa}^{(0)}_\e
-i\omega B_b (\bb{\alpha}^{(0)}_\delta + {\bb{\alpha}}^{*(0)}_\delta) + B_b^2 \bb{\sigma}^{(0)}_\e  
= c_b - B_b\rho \;.
\end{split}
\end{equation}
This identity is of the same form as \eqref{BHallWI}, with the modification $B \to B_b \equiv B + b $ contributing 
to the charge response, and $c \to c_b \equiv c + c_{el}$ incorporating an additional contribution to the thermal response
from the conduction electrons $c_{el}$, without changing the topological charge density.
The quantities $c$ and $b$ are constant and independent of $B$, 
while $c_{el}$ (also measurable) is expected to be proportional to $B$ and 
depends on the strength of the binding. $b, c, c_{el}$ are expected to be readily identifiable experimentally. 
In particular, $b$ can be identified from a step-function-like signature in 
the Hall conductivity $\bb{\sigma}_\e$ \cite{HC1,HC2,HC3,HC4}, as one passes into and out of a phase in which Skyrmions develop a finite density $c$.
Such behavior will also confirm the presence of a nonzero density $c$, 
which will likewise produce a similar step-function-like contribution in 
the thermal Hall conductivity $\bb{\kappa}_\e$ with an additional $B$-dependent $c_{el}$, 
by sweeping the magnetic field $B$ or the temperature $T$ independently.  

In the absence of ferromagnetic binding between the Skyrmion and conduction electron spins, the electric Hall conductivity would only pick up contributions from the conduction electrons, 
and $B_b$ would reduce to $B$. On the other hand, the thermal Hall conductivity would include  both contributions, 
$c$ and $c_{el}$, with the latter being independent of $B$. 

A second, alternative way to incorporate the interaction between thermal and charged responses is to impose the following operator relation 
\begin{equation}\label{TJ}
T^{0i} = \mu J^i \;,
\end{equation}
where $\mu$ parameterizes 
the strength of the coupling between the spins of the Skyrmion and the conduction electron. 
Then the momentum transport is tied to the charge transport as 
$\bb{\kappa} = \mu \bb{\alpha} = \mu {\bb{\alpha}}^* = \mu^2 \bb{\sigma}$.%
\footnote{Similar relations to \eqref{TJ} have been used to impose Galilean invariance in related systems
\cite{Read:2008rn,Read:2010epa,Hoyos:2011ez,Bradlyn:2012ea,Geracie:2014nka,Hoyos:2014lla,Hoyos:2015yna}.} 
The relation \eqref{TJ} implies a distinct experimental signature.  
At zero momentum, the analogue of \eqref{bBHallWI} becomes 
\begin{equation}
\begin{split}
&\bb{\sigma}^{(0)}_\delta =  -\frac{i\omega_c}{\omega} \frac{c -B\rho}{\mu^2(\omega^2-\omega_c^2)}\;, ~~
\bb{\sigma}^{(0)}_\e = \frac{c - B\rho}{\mu^2(\omega^2-\omega_c^2)}\;,
\end{split}
\end{equation} 
where $ \omega_c = \frac{B}{\mu}$.
For small magnetic field $ \omega_c \ll \omega$, the Hall conductivity is directly related to the 
topological charge density $ \bb{\sigma}^{(0)}_\e \approx \frac{c}{\mu^2 \omega^2}$ and 
$\bb{\sigma}^{(0)}_\delta \approx 0 $. In the opposite limit with large magnetic field $ \omega_c \gg \omega$,
$ \bb{\sigma}^{(0)}_\e \approx 0$ and 
$\bb{\sigma}^{(0)}_\delta \approx  \frac{-i\rho}{\mu\omega} $.
Such behavior can easily be measured.    
It would be interesting to find a material with Skyrmions that displays these properties. 

The momentum independent Ward identities are the same as \eqref{BHallWI} whether or not 
the system has translation invariance. 
If the system of interest has translation symmetry, one can use \eqref{BWIHallViscosity} with 
the modification $B \to B_b$ for momentum dependent Hall transport measurements. 

In the presence of translation symmetry, it is also simple to measure the Hall viscosity similarly to eq. \eqref{MeasuringHV}. Dividing eq. \eqref{BWIHallViscosity} by the second equation of \eqref{BHallWI} with $ B \to B_b, c \to c_b$, and taking the approximation $\omega/B_b \to 0$ and the limit $q^2 \to 0$, we obtain 
\begin{equation} \label{MeasuringHV2}
\begin{split}
\eta_H = (c_b - B_b \rho)  \frac{ \bb{\sigma}_\epsilon^{(2)} }{\bb{\sigma}^{(0)}_\epsilon} \;.
\end{split}
\end{equation} 
In the opposite limit $B_b/\omega \to 0 $, $\eta_H$ reduces to eq. \eqref{MeasuringHV} with the modification $c \to c_b - B_b \rho $. Note that this identification of $\eta_H$ can also be applied to systems without Skyrmions, such as quantum Hall systems. 

{\it Ward identities for conductors without translation invariance:} 
If spatial translation symmetry is broken (still assuming time-translation and rotation invariance), more physical quantities can come into play. 
In particular, the Ward identity is given by eq. \eqref{BWIHallViscosity} with the replacement 
$\eta_H \to  \eta_H+ \frac{\ell}{2}$ as in the neutral case 
\footnote{Likewise, the magnetization $M$, from the expectation value of the current 
$
\langle J^i \rangle = \epsilon^{ik} \partial_k M ,
$
can contribute in the combination $ \zeta -\frac{i}{\omega} \left( p - B M\right)$, 
replacing the bulk viscosity $\zeta$ \cite{Hoyos:2015yna}.}.
This Ward identity directly relates conductivities, angular momentum and Hall viscosity.
Recent experiments on metallic MnSi have studied transport properties and angular momentum 
\cite{SpinTorque1,SpinTorque2}. 
While Hall viscosity has not previously been discussed in the context of Skyrmion physics, 
it might play an important role and has a chance to be observed for the first time in active 
experiments.

{\it Acknowledgements:} 
We thank Sumit Das, Lance De Long, Michael Eides, 
Ori Ganor, Petr Ho\v{r}ava, Carlos Hoyos, Seungjoon Hyun, Seok Kim, Elias Kiritsis, 
Kimyeong Lee, Keh-Fei Liu, Hitoshi Murayama, Yaron Oz, Sumiran Pujari, Shinsei Ryu, 
Sang-Jin Sin, David Tong, Oskar Vafek, and Piljin Yi 
for helpful discussions. We are especially thankful to Jung Hoon Han 
for numerous illuminating discussions on Skyrmions and invaluable comments on the draft. 
Finally we would like to thank our referees for numerous helpful comments. 
BSK is grateful to the members of the Berkeley Center for Theoretical Physics, Berkeley, and 
KIAS, Seoul, for their warm hospitality during his visits. 
This work is partially supported by NSF Grant PHY-1214341.

\appendix
\onecolumngrid

\newpage

\section*{Supplemental Material for ``Skyrmions and Hall Transport''}

Here we drive the general Ward identity, equation (14) of our Letter,
in the presence of a uniform external magnetic field $B$ with conserved current $J^\mu$. 
The momentum density and its conservation equation are modified as in equations (12) and (13). 
Taking two time derivatives of the retarded Green function 
$G^{0i,0j}(x^\mu; {x'}^\mu)=i\theta(x^0\-{x'}^0)\vev{[T^{0i}(x^\mu),T^{0j}({x'}^\mu)]}$ gives 
\begin{align}\label{BStep1}\tag{S1}
\begin{split}
\partial_0 \partial'_{0} G^{0i,0j} &= 
\d_n\d_m'  G^{nj,ml}+B\e^j_{\ n}\d_0' G^{n,0l}+B\e^l_{\ m}\d_0 G^{0j,m}-B^2\e^j_{\ n}\e^l_{\ m} G^{n,m} \\
&-\frac{1}{2}(\d_0-\d_0')\left(\delta(x^0\- {x'}^0) C^{0j,0l}\right) \;,
\end{split}
\end{align}
where $G^{a,b}(x^\mu; x'^\mu)=i\theta(x^0\- {x'}^0)\vev{[V^{a}(x^\mu),V^{b}( x'^\mu)]}$ are the 
retarded correlators. The index $ a$ can either represent one or two indices, depending on context: e.g., $V^a$ can represent either the momentum density components $T^{0i}$ or a current $J^i$. Note that the first line in \eqref{BStep1} is a consequence of the conservation equation 
$\partial_\mu T^{\mu i} = B \epsilon^i_j J^j$, 
while the second line comes from a contact term with an equal time correlator
$C^{a,b}(x^0;\vec x, \vec x')=i \langle [V^{a}(x^0,\vec x),V^{b}(x^0, \vec x')] \rangle$. 
There are four other terms 
$$ \frac{1}{2}\delta(x^0\- x'^0) 
\big[ \d_n C^{nj,0l}-\d_m' C^{0j,ml} + B \left(\e^j_{\ n}C^{n,0l}\-\e^l_{\ m}C^{0j,m}\right) \big]$$ 
that vanish in the presence of spacetime translation and rotation symmetries. 
See Ref. [11]  for a discussion of the general case. 

Using time translation symmetry, we perform a Fourier transform $\int d(x^0\- x'^0) e^{iq_0 (x^0\- x'^0)}$ on \eqref{BStep1}, followed by 
$$G^{a,b}(x^0\- x'^0; \vec x , \vec x') = \int  \frac{dq_0}{2\pi} 
e^{-i q_0 (x^0\- x'^0) } \hat{G}^{a,b}(q_0; \vec x , \vec x') $$ using  
equation (3)  and  
$i[T_B^{0i}(\vec x), \mathcal O(\vec x')]=\d_i\mathcal O(\vec x)\delta^{2}(\vec x - \vec x')$, 
where $T_B$ is the momentum generator in the presence of the magnetic field given in equation (12).  
We evaluate the contact term in \eqref{BStep1} as 
$i\omega  C^{0j,0l} =-i\omega  \e^{jl} \big[\e^{n}_{\ m}\d_n\vev{T^{0m}} 
+c - B \rho \big] \delta^{2}(\vec x - \vec x')$, where $ q_0 \equiv \omega$.
The first term is incompatible with translation and rotation symmetries, 
and we discard it. 
$c$ and $\rho= \vev{J^0}$ are the topological number density and charge density. 
The Ward identity becomes  
\begin{align}\label{MiddleStep}\tag{S2}
\begin{split}
\omega^2 \hat{G}^{0j,0l} &-i\omega B\e^j_{\ n}\hat G^{n,0l}
+i\omega B\e^l_{\ m}\hat G^{0j,m}
+ B^2\e^j_{\ n}\e^l_{\ m} \hat G^{n,m} \\
&=\d_n\d_m' \hat G^{nj,ml}  \-i\omega \e^{jl} ( c  - B \rho) \delta^{2}(\vec x \- \vec x') \;. 
\end{split}
\end{align}

Using space translation symmetry, we perform a Fourier transform 
$\hat G^{a,b}(\omega;\vec x \- \vec x')=\int \frac{d^2 q}{(2\pi)^2} 
e^{i\vec{q}\cdot(\vec{x}\-{\vec{x}'})}\widetilde {G}^{a,b}(\omega,\vec{q}) $. 
To extract useful information, we exploit rotational invariance to 
rewrite the retarded Green's functions in the form 
\begin{equation}\tag{S3}
\begin{split}
\widetilde G^{a,b}&= \Pi_\delta \delta^{ab}+\Pi_\e  \e^{ab}+q^a q^b\Pi_{q}
+(\e^{bo}q^a+\e^{ao}q^b)q_o\Pi_{q\e} \;, \\
\widetilde G^{ij,kl} &= -i\omega \Big[ {\eta}\left(\delta^{ik}\delta^{jl}+\delta^{il}\delta^{jk}
-\delta^{ij}\delta^{kl} \right)+{\zeta}\delta^{ij}\delta^{kl}  
+\frac{{\eta}_H}{2}(\e^{ik}\delta^{jl}+\e^{il}\delta^{jk}+\e^{jk}\delta^{il}+\e^{jl}\delta^{ik}) \Big] \;. 
\end{split}
\end{equation}
Here $\Pi^{TT}, \Pi^{TJ}, \Pi^{JT}, \Pi^{JJ}$ are four different types of form factors,    
and $\eta, \zeta, \eta_H $ are the shear, bulk and Hall viscosities, respectively. 
Plugging them into \eqref{MiddleStep} and doing some algebra, we arrive at 
\begin{equation}\label{BMomWIFull}\tag{S4}
\begin{split}
& \delta^{jl}  \Big[\omega^2 \Pi^{TT}_\delta 
\+i\omega B (\Pi^{JT}_\e \+ \Pi^{TJ}_\e) \+ B^2 \Pi^{JJ}_\delta \+ B^2 q^2 \Pi^{JJ}_{q} 
\- i\omega B q^2 (\Pi^{JT}_{q\e} \-\Pi^{TJ}_{q\e}) \Big] \\
&\+\e^{jl} \Big[\omega^2 \Pi^{TT}_\e
\-i\omega B (\Pi^{JT}_\delta \+ \Pi^{TJ}_\delta) \+ B^2 \Pi^{JJ}_\e \- 
\frac{i\omega B}{2} q^2 (\Pi^{JT}_{q} \+\Pi^{TJ}_{q}) \Big] \\
&\+ q^j q^l  \Big[ 
\omega^2 \Pi^{TT}_{q} \+ 2i\omega B (\Pi^{JT}_{q\e} \- \Pi^{TJ}_{q\e}) \- B^2 \Pi^{JJ}_{q} \Big]  
\+ (\e^{jo}q^l \+ \e^{lo}q^j) q_o \Big[ 
\omega^2 \Pi^{TT}_{q\e} \-\frac{i\omega B}{2} (\Pi^{JT}_{q} \- \Pi^{TJ}_{q}) \- B^2 \Pi^{JJ}_{q\e} \Big] \\
&= \e^{jl} i \omega \big[-c + B\rho - q^2 \eta_H \big]  -
i\omega \delta^{jl} q^2  \eta -  i\omega q^j q^l  \zeta   \;.
\end{split}
\end{equation}
We identify $\Pi^{TT} = -i\omega \bb{\kappa}, \Pi^{TJ} = -i\omega \bb{\alpha}, \Pi^{JJ} = -i\omega \bb{\sigma}$, 
where $\bb{\kappa},\bb{\alpha}, \bb{\sigma} $ are thermal, thermoelectric, and electric conductivity tensors. 
Matching tensor structures in \eqref{BMomWIFull}, we find four independent Ward identities. 
The resulting equation appears as equation (15).

\end{document}